# Superconductivity in the A15-type $V_3(Os_{1-2x}Si_xGe_x)$ medium-entropy alloys

Yucheng Li[1,#], Kuan Li[1,#], Lingyong Zeng[1], Rui Chen[1], Jingjun Qin[1], Shuangyue Wang[1], Huixia Luo[1,2,3,4*]

[1] School of Materials Science and Engineering, Sun Yat-sen University, No. 135, Xingang Xi Road, Guangzhou, 510275, P. R. China
[2] State Key Laboratory of Optoelectronic Materials and Technologies, Sun Yat-sen University, No. 135, Xingang Xi Road, Guangzhou, 510275, P. R. China
[3] Key Lab of Polymer Composite & Functional Materials, Sun Yat-sen University, No. 135, Xingang Xi Road, Guangzhou, 510275, P. R. China
[4] Guangdong Provincial Key Laboratory of Magnetoelectric Physics and Devices, Sun Yat-sen University, No. 135, Xingang Xi Road, Guangzhou, 510275, P. R. China
[#] These authors contributed equally to this work.
[*] Corresponding author (email: luohx7@mail.sysu.edu.cn)

**ABSTRACT**   Cubic A15-type superconducting alloys continue to fascinate the academic and industrial fields because they mainly support the largest market for low-temperature superconducting applications and show exotic physical properties. Medium-/high-entropy alloys (MEAs-HEAs) can be employed stably under extreme conditions due to their high mechanical hardness and excellent irradiation tolerance. Combining with the features of the A15-type superconductor and MEAs-HEAs, we design a series of previously unreported A15-type $V_3(Os_{1-2x}Si_xGe_x)$ ($x$ = 0.333, 0.375, 0.425) MEA superconductors, which can be obtained by an arc melting method. Resistivity, magnetic susceptibility, and specific heat measurements indicate that all of them are type-II bulk superconductors. The superconducting transition temperature ($T_c$) exhibits an upward trend with the systematic reduction of Os concentration. Additionally, the upper critical field of the $V_3(Os_{0.333}Si_{0.333}Ge_{0.333})$ sample is larger than the Pauli limit, suggesting it may be robust against magnetic fields due to spin-orbit coupling induced by the heavy Os atoms. These findings not only advance our understanding of emergent phenomena in entropy-stabilized A15-type alloys but also expand the members of new superconductors.

**INTRODUCTION**

Superconductors with the cubic A15 structure, which were discovered mostly between the 1950s and 1970s, are representative of metal-based superconductors [1-5]. Until the copper-oxide superconductors were discovered in 1986, the A15 structure held the record highest superconducting transition temperature ($T_c$), around 23 K for Nb$_3$Ge [6, 7], which has since been commercialized. Another classical A15-type superconductor, V$_3$Si, with the isostructure of Nb$_3$Ge ($T_c$ ~ 17 K, valence electron count (VEC) = 4.7[8]), demonstrates many charming features, even though its critical temperature is surpassed by Nb$_3$Ge. Recent observations have confirmed that V$_3$Si exhibits many unusual characteristics, including anisotropy in the upper critical field ($H_{c2}$) and specific heat, the presence of the de Haas-van Alphen effect, a phase transition above $T_c$, soft acoustic phonons, and a large ratio of $T_c$ to the Fermi temperature, four- and eightfold degenerate fermions [9-15]. However, another isostucture A15-type superconductor V$_3$Ge is considered to be a conventional type-II superconductor with a $T_c$ of 6.1 K [16, 17].

More recently, medium/high-entropy materials (MEMs-HEMs) with multiple elements mixed in equimolar or near-equimolar ratios have emerged as a new class of functional material due to their distinct mechanical hardness and rich chemical/physical properties [18-21]. Research on superconductivity of MEMs-HEMs started relatively late, starting in 2014 in the high-entropy alloy (HEA) Ta-Nb-Hf-Zr-Ti, yet it subsequently attracted considerable interest in their pairing mechanism [22]. Since it is generally considered that high disorders in crystalline superconductors generally are not conducive to the formation of Cooper pairs, this is partially attributed to the decrease of the density of states (DOSs) or the increase of effective Coulomb repulsion between paired electrons [23, 24]. In addition, many unique physical properties(e.g., topological bands, against high pressure, strong electron-phonon coupling, the coexistence of magnetism and superconductivity, and exceptionally high upper critical fields) have been found in these complex superconducting MEMs-HEMs [25-31]. So far, the currently known MEM-HEM superconductors adopt various crystal structures, such as body-centered cubic (BCC) structure, face-centred cubic (FCC) structure, hexagonal close-packed structure (HCP) structure, σ structure, CsCl structure, and complex $\alpha/\beta$ -Mn structure[32, 33]. Meanwhile, investigations into medium-entropy/high-entropy alloy (MEA-HEA) superconductors possessing the A15 structure are underway. Nakahira Yuki, et al.[17] applied the compositionally complex alloy (CCA) concept to an A15-type superconductor by synthesizing V$_3$X (X = Al, Si, Ga, Ge, Sn), which achieved a ~50 % higher $H_{c2}(0)$ while maintaining a comparable transition temperature ($T_c \approx 6.3$ K) compared to binary $V_3Ge$. The study by Aichi Yamashita et al.[34] demonstrated the successful synthesis of A15-type HEA superconductors in the $Nb_3(Al,Sn,Ge,Ga,Si)$ system, in which ($Nb_3Al_{0.2}Sn_{0.2}Ge_{0.2}Ga_{0.2}Si_{0.2}$) and ($Nb_3Al_{0.3}Sn_{0.3}Ge_{0.2}Ga_{0.1}Si_{0.1}$) show $T_c$ of 9.0 K and 11.0 K, respectively. Liu et al. reported a series of A15-type $V_{5+2x}Nb_{35-x}Mo_{35-x}Ir_{10}Pt_{15}$ ($0 \leq x \leq 10$) HEAs, and they found $T_c$s and zero-temperature upper critical fields Hc2(0)s both decrease monotonically with the increase of V content $x$, while $T_c$ increases as the valence electron concentration (VEC) increases [35]. Moreover, the phase transformation in A15-type MEAs-HEAs can be effectively tailored through doping and annealing strategies. Wu, et al.[36] reported the first observation of temperature-driven polymorphism and its direct impact on superconductivity in $(V_{0.5}Nb_{0.5})_{3-x}Mo_xAl_{0.5}Ga_{0.5}$ HEAs, in which annealing transforms the non-superconducting bcc-type $(V_{0.5}Nb_{0.5})_{3-x}Mo_xAl_{0.5}Ga_{0.5}$ into an A15 structure, and the A15-type polymorph with $x = 0.2$ achieved a record-high $T_c$ of 10.2 K and $H_{c2}$ of 20.1 T among all known MEA-HEA superconductors under atmospheric pressure.

Previous works have established the profound influence of composition on the superconducting and structural properties of A15-type MEAs-HEAs, yet the exploration of new compositional spaces remains crucial for a comprehensive understanding and performance optimization. As a heavy 5$d$ transition metal, Os is strategically introduced to incorporate strong SOC, which provides a unique platform to investigate the enhancement of the upper critical field in A15-type MEAs. Furthermore, the distinct difference in VEC between Os (group 8) and Si/Ge (group 14) offers a broad compositional window to precisely tune the average VEC of the system, enabling the systematic modulation of $T_c$ as guided by the Matthias empirical rule. Combining with the exotic properties of the A15-type superconductors and MEAs-HEAs, we design and synthesize a series of previously unreported A15-type $V_3(Os_{1-2x}Si_xGe_x)$ ($x = 0.333, 0.375, 0.425$) MEA superconductors with the VEC of 5.08, 5, and 4.9, respectively, which can be obtained by an arc melting method. Resistivity measurements demonstrate that they are type-II superconductors, with a $T_c$ of 4.48 K, 4.73 K, and 5.62 K, respectively. $T_c$ decreases monotonically with the increase of VEC and Os content. An upper critical field of $V_3(Os_{0.333}Si_{0.333}Ge_{0.333})$ is 8.97 T, which exceeds the Pauli limiting field, based on the Ginzburg-Landau (GL) model. Furthermore, specific heat measurement results indicate those are $s$-wave superconductors, with the nature of bulk superconductivity. Furthermore, the zero-field critical current density ($J_c$) values for all $V_3Os_{1-2x}Si_xGe_x$ ($x = 0.333, 0.375, 0.425$) MEAs are evaluated to be in the range of 1.48 - 8.48 × $10^6$ A/cm$^2$ at 2 K. These values significantly surpass the recognized practical benchmark of $10^5$ A/cm$^2$ required for superconducting devices, with the $x = 0.375$ sample demonstrating superior current-carrying capability, highlighting the promise of these alloys for large-scale applications.

**EXPERIMENTAL SECTION**

The polycrystalline samples of $V_3(Os_{1-2x}Si_xGe_x)$ ($x$ = 0.333, 0.375, 0.425) were synthesized utilizing a simple arc-melting method. Initially, stoichiometric amounts of vanadium (V, 99.9%, Aladdin), osmium (Os, 99.95%, Adamas-beta), silicon (Si, 99.99%, Macklin), and germanium (Ge, 99.999%, Macklin) powders were meticulously blended through thorough milling to ensure homogeneity. This mixture was then pressed into a cylindrical form, transferred to an electric arc furnace, purged with argon three times to remove air, and finally melted in an argon atmosphere at 0.5 atmospheres. The resulting compound was then homogenized by melting the ingot four times. The weight loss during the melting process was minimal and deemed negligible.

To analyze the phase composition and lattice structure of the synthesized samples, powder X-ray diffraction (XRD) measurements were collected using a Rigaku MiniFlex (Cu K$\alpha$1 radiation) at a constant scan speed of 1 °/min from 20° to 100°. The collected XRD data were subsequently refined using the Rietveld method, providing detailed insights into the crystal structure. Concurrently, the microstructural characteristics of the samples were examined via scanning electron microscopy (SEM, EVO MA10) coupled with energy dispersive spectroscopy (EDS), enabling the assessment of elemental composition and morphology. Furthermore, the temperature-dependent resistivity was measured using a four-probe technique, while the field-cooled (FC), zero-field-cooled (ZFC) magnetization, specific heat capacity, and critical current density ($J_c$) were determined employing a physical property measurement system (PPMS, DynaCool, Quantum Design, Inc.).

**RESULTS**

X-ray diffraction (XRD) patterns of the A15-type MEA superconducting materials $V_3(Os_{1-2x}Si_xGe_x)$ ($x$ = 0.333, 0.375, 0.425) were collected over a wide angle range (2θ from 20° to 90°) as shown in Fig. 1a. The obtained XRD data were fitted using the Rietveld refinement method. The refined XRD patterns show that the experimental data are in good agreement with the calculated data, as reflected by the low value of the goodness of fit parameter $\chi^2$. The peak positions in the XRD pattern are accurately assigned to the corresponding crystal planes of the A15-type structure. The high-intensity peaks at 37°, 42°, and 46° are assigned to the (200), (210), and (211) planes, which are characteristic reflections of the A15-type crystal structure.

In the A15-type structure ($A_3B$), the 6$c$ sites are typically occupied by transition metal chains (V in our case), while the 2$a$ sites are occupied by the B-group elements[37-39]. Figure S1(a-c) shows the Rietveld refinement profiles of the XRD for $V_3Os_{1-2x}Si_xGe_x$ ($x$ = 0.333, 0.375, 0.425) samples, (d) shows the crystal structure of the $V_3Os_{1-2x}Si_xGe_x$ ($x$ = 0.333) sample. The refinement was conducted by assigning V to the *6c* sites and randomly distributing Os, Si, and Ge at the *2a* sites. This model yielded reliable factors of $R_{wp}$ = 11.78%, $R_p$ = 8.11%, $R_e$ = 9.61% and $\chi^2$ = 1.5023 for $x$ = 0.333 sample. The factors of other samples are shown in Table S1. Given the inherent lattice distortion and chemical complexity of MEAs, these values represent a high-quality fit and confirm that the A15-type ordered structure is well-maintained.

Furthermore, according to the classic Labbé–Friedel model and experimental studies by Sweedler et al. [37], superconductivity in A15-type alloys is highly sensitive to atomic ordering. Any significant antisite disorder would interrupt the one-dimensional continuity of V-3d states, resulting in a drastic suppression of the density of states at the Fermi level and consequently a collapse of $T_c$. The observation that our samples maintain a stable $T_c$ and exhibit high $J_c$ provides strong support for our Rietveld refinement results, which indicate that Os preferentially occupies the 2a sites. Fig. 1b illustrates the variation of lattice parameters and unit cell volume as a function of $x$ content for the $V_3(Os_{1-2x}Si_xGe_x)$ ($x$ = 0.333, 0.375, 0.425) series. Both the lattice parameter and the unit cell volume exhibit a monotonic decrease with increasing $x$ content. This contraction of the crystal lattice can be attributed to the substitution of Os atoms by Si and Ge atoms; specifically, the atomic radius of Si (111 pm) and Ge (122 pm) are significantly smaller than that of Os (135 pm) and V (134 pm), leading to a reduction in the overall lattice dimensions as the doping level $x$ increases.

SEM-EDS is further used to characterize the surface morphology and element distribution of the sample of $V_3(Os_{1-2x}Si_xGe_x)$ ($x$ = 0.333, 0.375, 0.425) superconducting material, as shown in Fig. 1c. The spectrum identified all the expected elements present in the sample. In addition, the element distribution of all $V_3(Os_{1-2x}Si_xGe_x)$ ($x$ = 0.333, 0.375, 0.425) superconducting samples was relatively uniform, indicating that the elements were well mixed during the synthesis process.

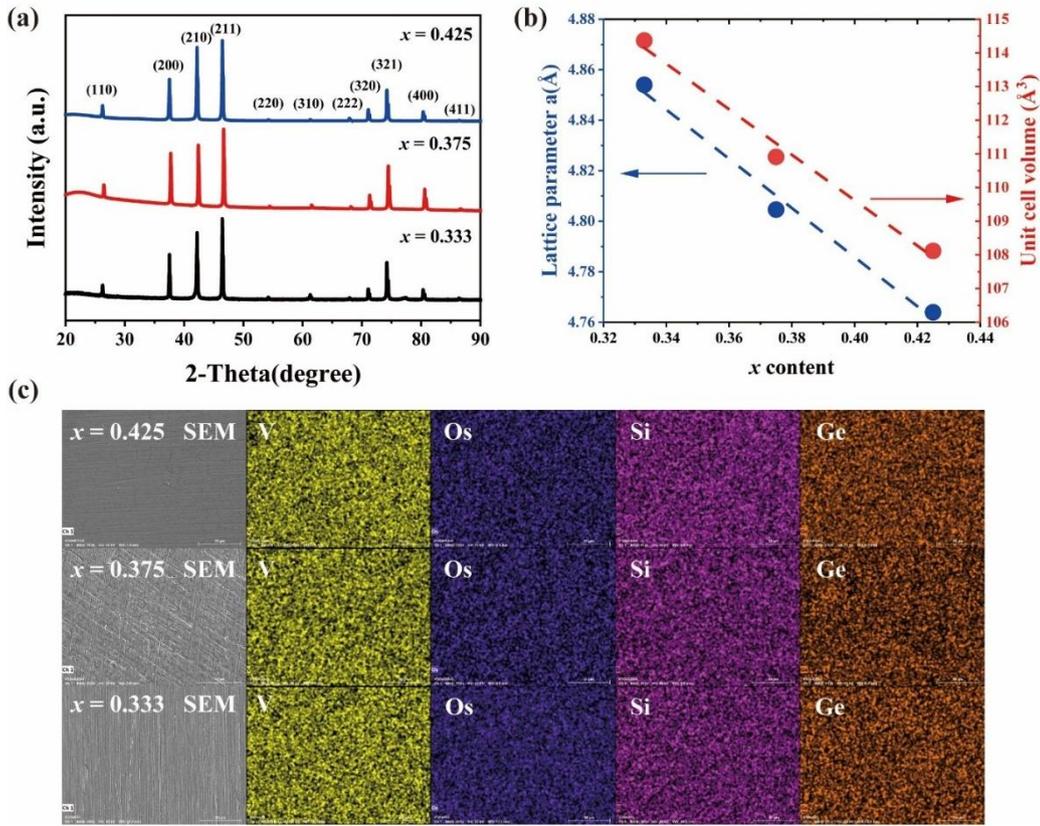

Figure 1 (a) XRD patterns of the $V_3Os_{1-2x}Si_xGe_x$ ($x$ = 0.333, 0.375, 0.425) alloys. (b) Lattice parameters and unit cell volume as a function of the fraction of $V_3Os_{1-2x}Si_xGe_x$ ($x$ = 0.333, 0.375, 0.425) MEAs. (c) SEM images and the corresponding EDS element mappings of $V_3Os_{1-2x}Si_xGe_x$ ($x$ = 0.333, 0.375, 0.425) MEAs.

Resistivity measurements of $V_3Os_{1-2x}Si_xGe_x$ ($x$ = 0.333, 0.375, 0.425) superconducting materials in zero magnetic field were performed in the temperature range of 1.8 K to 300 K, as shown in Fig. 2a. As the temperature decreases, the resistance of the sample initially shows typical metallic properties, with the resistance gradually decreasing. This is consistent with the general trend of metals, where the scattering of electrons by lattice vibrations (phonons) and defects decreases with decreasing temperature. The residual resistivity ratio (RRR) of each sample was calculated as RRR = $\rho(300\ K)/\rho(10\ K)$, which is used to evaluate the purity and conductivity. Here, RRR = 1.09, 1.05 and 1.13 for $V_3(Os_{0.333}Si_{0.333}Ge_{0.333})$, $V_3(Os_{0.250}Si_{0.375}Ge_{0.375})$ and $V_3(Os_{0.150}Si_{0.425}Ge_{0.425})$, respectively. Fig. 2b shows the temperature-dependent electrical resistivity ($\rho(T)$) of $V_3Os_{1-2x}Si_xGe_x$ ($x$ = 0.333, 0.375, 0.425) over 1.6 - 8 K. The $T_c^{onset}$ = 4.89 K, 5.25 K, and 5.8 K, respectively, is determined as the resistance begins to deviate from the linear behavior of the normal state. The zero resistance state occurred at $T_c^{zero}$ = 4.03 K, 4.5 K, and 5.25 K, respectively, where the resistance completely disappears, and the superconducting state is established. Based on the $\rho(T)$ data, the $T_c^{mid}$ is determined to be 4.48 K, 4.73 K, and 5.62 K, respectively, where $T_c$ is defined as the 50% drop in resistivity relative to the normal state value. The superconducting transition width of samples (defined as the temperature range over which the resistance drops from 90% to 10% of the normal state value) is relatively narrow at 0.86 K, 0.75 K, and 0.55 K. Based on the aforementioned data, we observe that $T_c$ increases with decreasing Os content in sample $V_3Os_{1-2x}Si_xGe_x$ ($x$ = 0.333, 0.375, 0.425) MEAs.

The superconducting behavior of $V_3Os_{1-2x}Si_xGe_x$ ($x$ = 0.333, 0.375, 0.425) materials is significantly affected when an external magnetic field is applied. As the magnetic field increases, the $T_c$ decreases. The magnetic field penetrates the superconductor in the form of quantized magnetic flux lines, destroying Cooper pairs and increasing the scattering of electrons, thereby increasing the resistance. We measured the resistivity curves at magnetic field strengths 0 - 7 T for $V_3Os_{1-2x}Si_xGe_x$ ($x$ = 0.333, 0.375, 0.425) as shown in Fig. 2c and Fig. S2. The $T_c$ decreases with the enhancement of perpendicular fields. The upper critical magnetic field ($\mu_0H_{c2}(0)$) is a key parameter that determines the maximum magnetic field that a superconductor can withstand while maintaining a superconducting state. To obtain experimental data on the relationship between the $T_c$ and the applied magnetic field, the Ginzburg-Landau (G-L) theory was used for further fitting, as shown in Fig. 2d: $\mu_0H_{c2}(T) = \mu_0H_{c2}(0) \times \frac{1-(T/T_c)^2}{1+(T/T_c)^2}$, yielding $\mu_0H_{c2}(0)$ = 8.97 T, 7.68 T, and 8.86 T for $V_3Os_{1-2x}Si_xGe_x$ ($x$ = 0.333, 0.375, 0.425), respectively. The orbital limiting field can be described by the Werthamer-Helfand-Hohenberg (WHH) model: $\mu_0H_{c2}(T) = -0.693T_c \frac{d\mu_0H_{c2}}{dT}\big|_{T=T_c}$, The slope $\frac{d\mu_0H_{c2}}{dT}\big|_{T=T_c}$ of $V_3Os_{1-2x}Si_xGe_x$ ($x$ = 0.333, 0.375, 0.425) can be obtained −2.248 T/K, −2.0 T/K, and −1.845 T/K, respectively, from the linear fitting. The calculated $\mu_0H_{c2}(0)$ of the WHH model is 6.98 T, 6.56 T, and 7.19 T, respectively. Based on BCS theory, the Pauli limiting field for a superconductor can be described as $\mu_0H_{c2}^p(0) = 1.86 \times T_c$, that is, $\mu_0H_{c2}^p(0)$ = 8.33 T, 8.80 T and 10.43 T for $V_3Os_{1-2x}Si_xGe_x$ ($x$ = 0.333, 0.375, 0.425), respectively. Notably, $V_3(Os_{0.333}Si_{0.333}Ge_{0.333})$ exhibits an upper critical magnetic field that exceeds the Pauli limit. This is likely attributed to strong spin-orbit scattering, which suppresses the Pauli paramagnetic pair-breaking effect. Fig. 2e shows the temperature-dependent magnetic susceptibility for $V_3Os_{1-2x}Si_xGe_x$ ($x$ = 0.333, 0.375, 0.425). The magnetization data (defined as $\chi_v = \frac{M_V}{H}$, where $H$ is the applied magnetic field, and $M_V$ is the volume magnetization) was measured during zero field cooling (ZFC) under an applied magnetic field of 30 Oe. At higher temperatures, the magnetic susceptibility of the ZFC curve is close to zero and remains essentially unchanged, indicating that the material is in a normal state and responds weakly to the magnetic field. As the temperature decreases, at about 4.25 K, 4.36 K, and 5.39 K, respectively, the magnetic susceptibility begins to decrease rapidly and eventually approaches −1, which is a typical feature of superconductors. At these temperatures, all $V_3Os_{1-2x}Si_xGe_x$ ($x$ = 0.333, 0.375, 0.425) samples enter the superconducting state, and the magnetic field is completely expelled from the body. Fig. S3 shows the magnetic isotherms measured in the temperature range of 1.8 - 2.6 K. In the case of perfect field response, $M_{fit} = a + bH$ is determined by linear fitting of the low-field region of the magnetization data at 1.8 K. Then, the demagnetization factor $N$ is obtained according to the formula $-a = 1/4\pi(1-N)$. According to the slope value ($a$), the $N$ value is determined to be 0.43. The $4\pi\chi_v(1-N)$ (ZFC) value is close to −1 at 1.8 K. The observation of a 100 % Meissner volume fraction across all specimens confirms their bulk superconductivity, with the superconducting shielding fraction reaching nearly 100 % once the demagnetizing factor N is taken into account. The $M$-$M_{fit}$ diagram in Fig. S3 was plotted using the low-field linear fit of the magnetization data. It should be noted that the value of 0.1 emu/cm$^3$ is less than 2 % of $M$ at 1.8 K, so the lower critical field $\mu_0H_{c1}(0)$ can be accurately calculated. The extracted points are fitted using the empirical formula $\mu_0H_{c1}(T) = \mu_0H_{c1}(0)(1-(T/T_c)^2)$, where $T_c$ is the superconducting transition temperature and $\mu_0H_{c1}(0)$ is the lower critical field at 0 K. Fig. 2f shows the temperature-dependent lower critical fields $\mu_0H_{c1}$ for $V_3Os_{1-2x}Si_xGe_x$ ($x$ = 0.333, 0.375, 0.425). For the $V_3Os_{1-2x}Si_xGe_x$ ($x$ = 0.333, 0.375, 0.425) samples, the estimated $\mu_0H_{c1}(0)$ = 8.36 mT, 7.13 mT, and 8.54 mT, respectively. The fitting line agrees well with the experimental data points, indicating that the adopted fitting method can well describe the relationship between the variation of the effective lower critical field and temperature. Notably, consistent with the behavior of the upper critical field, the $x$ = 0.333 sample exhibits a larger lower critical field compared to the $x$ = 0.375 sample, despite having a relatively lower $T_c$.

Based on the results of $\mu_0H_{c2}(0)$ and $\mu_0H_{c1}(0)$, we can calculate and extract various superconducting parameters for the $V_3Os_{1-2x}Si_xGe_x$ ($x$ = 0.333, 0.375, 0.425) MEAs. The GL coherence length, $\xi_{GL}(0)$ = 61 Å, 66 Å, and 61 Å, respectively, can be derived via the formula $\xi_{GL}^2(0) = \frac{\phi_0}{2\pi\mu_0H_{c2}(0)}$, where $\phi_0$ = 2.07×10$^{-15}$ Wb denotes the magnetic flux quantum. The magnetic penetration depth was derived from: $\mu_0H_{c1}(0) = \frac{\phi_0}{4\pi\lambda_{GL}^2(0)} \ln\frac{\lambda_{GL}(0)}{\xi_{GL}(0)}$, yielding $\lambda_{GL}(0)$ = 2685 Å, 2921 Å, and 2666 Å, respectively. Accordingly, from the formula, $\kappa_{GL}(0) = \lambda_{GL}(0)/\xi_{GL}(0)$, we obtain the GL parameter $\kappa_{GL}(0)$ = 44.3, 44.6, and 43.7, respectively. These value significantly exceeds the value $\frac{1}{\sqrt{2}}$, affirming that $V_3Os_{1-}$ $_{2x}Si_xGe_x$ ($x$ = 0.333, 0.375, 0.425) MEAs are type-II superconductors.

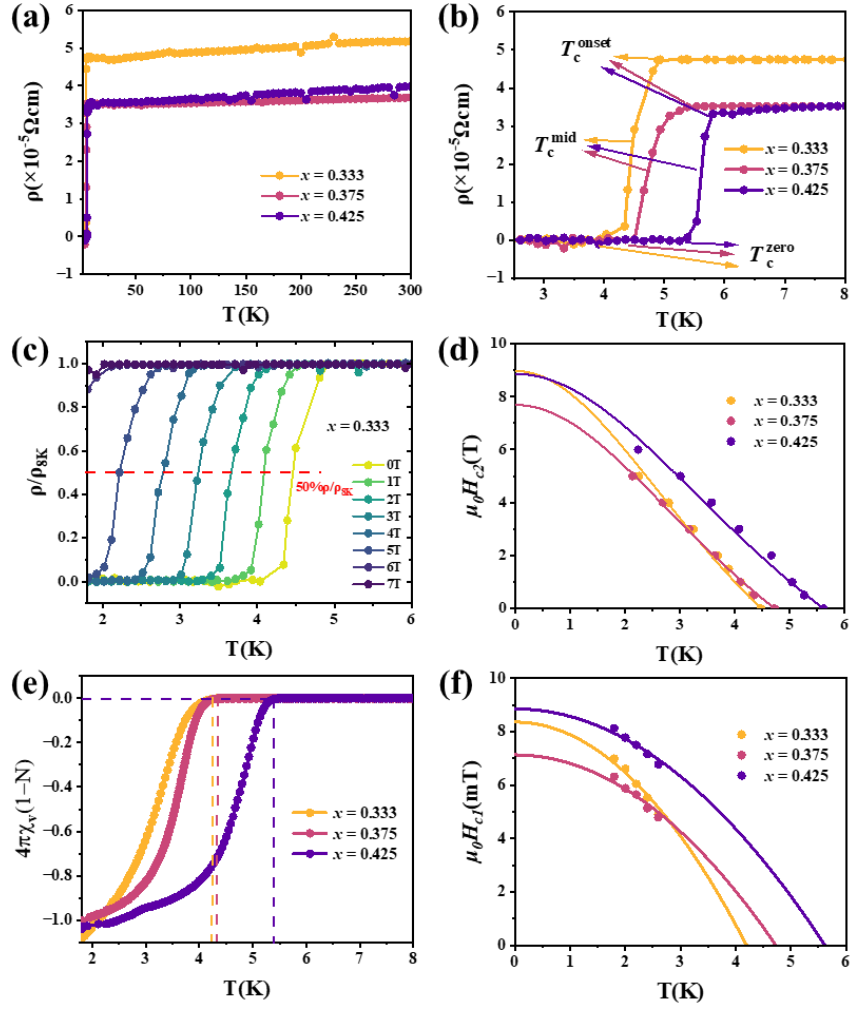

Figure 2 (a) The temperature-dependent resistivity of the $V_3OS_{1-2x}Si_xGe_x$ ($x$ = 0.333, 0.375, 0.425) MEA superconductor from 1.6 to 300 K. (b) The temperature-dependent resistivity from 1.6 to 8 K. (c) The resistivity transition at the field of 0-7 T for $V_3Os_{0.333}Si_{0.333}Ge_{0.333}$. (d) The temperature-dependent upper critical field $\mu_0H_{c2}$ with the Ginzburg-Landau function fitting. (e) The temperature-dependent magnetic susceptibility for $V_3Os_{1-2x}Si_xGe_x$ ($x$ = 0.333, 0.375, 0.425). (f) The temperature-dependent lower critical fields $\mu_0H_{c1}$ for $V_3Os_{1-2x}Si_xGe_x$ ($x$ = 0.333, 0.375, 0.425).

We performed the specific heat measurements on the $V_3Os_{1-2x}Si_xGe_x$ ($x$ = 0.333, 0.375, 0.425) samples to further verify the bulk superconductivities. We fit the specific heat data using the formula $C_p/T = \gamma + \beta T^2$, where $\gamma$ is the normal-state electronic coefficient, and β is the phonon-specific heat coefficient. Our fit gives $\gamma$ = 25.017 mJ·mol$^{-1}$·K$^{-2}$, $\beta$ = 0.080 mJ·mol$^{-1}$·K$^{-4}$ for $x$ = 0.425, $\gamma$ = 22.328 mJ·mol$^{-1}$·K$^{-2}$, $\beta$ = 0.089 mJ·mol$^{-1}$·K$^{-4}$ for $x$ = 0.375 and $\gamma$ = 19.445 mJ·mol$^{-1}$·K$^{-2}$, $\beta$ = 0.106 mJ·mol$^{-}$

$^1 \cdot K^{-4}$ for $x = 0.333$. The electronic specific heat, normalized as $C_{el}/\gamma T_c$, was derived by isolating the phonon component $\beta T^2$ from the raw experimental data $C_p$. Fig. 3a-c show the normalized electronic specific heat $C_{el}/\gamma T_c$ as a function of reduced temperature $T/T_c$ for $x = 0.425$, 0.375, and 0.333, respectively. The insets show the measured specific heat $C_p/T$ versus $T^2$. According to the $\alpha$ model[40], $C_{el}(T) = A \exp(-\Delta_0/k_B T)$, where $\Delta_0$ and $k_B$ are the superconducting gap at 0 K and Boltzmann constant, respectively, the normalized heat capacity jumps $\Delta C_{el}/\gamma T_c$ for the $x = 0.425$ sample are determined to be 1.35 at $T_c^{mid}$ and 1.22 at $T_c^{zero}$. Values of other samples are shown in Table 1. $\Delta C_{el}/\gamma T_c$ values of all samples are smaller than 1.43 according to the BCS theory, indicating the weak coupling. The $\alpha = \Delta_0/k_B T_c$ can be obtained by the function $\Delta C_{el}/\gamma T_c = (\Delta C_{el}/\gamma T_c)_{BCS}(\alpha/\alpha_{BCS})^2$. Here, $(\Delta C_{el}/\gamma T_c)_{BCS} = 1.43$, $\alpha_{BCS} = 1.76$. We obtained that coupling strength $2\Delta_0/k_B T_c = 3.26$, 3.30, and 3.43 for $x = 0.333$, 0.375, and 0.425, respectively. Again, these values are smaller than the weak-coupling BCS value (3.52).

The Debye temperature of the $V_3Os_{1-2x}Si_xGe_x$ ($x = 0.333, 0.375, 0.425$) alloys can be calculated by the following equation: $\Theta_D = (12\pi 4nR/5\beta)^{1/3}$, where n is the number of atoms per unit, which is 4 for $V_3Os_{1-2x}Si_xGe_x$ ($x = 0.333, 0.375, 0.425$) alloys, and R = 8.31 Jmol$^{-1}$K$^{-1}$ is the gas constant. By calculation, we get the Debye temperature $\Theta_D = 459.8$ K, 443.7 K, and 418.6 K for $x = 0.333$, 0.375, and 0.425. According to McMillan's theory, the electroacoustic coupling strength can be calculated by the following formula: $\lambda_{ep} = \dfrac{1.04 + \mu^* \ln(\frac{\Theta_D}{1.45 T_c})}{(1-0.62\mu^*)\ln(\frac{\Theta_D}{1.45 T_c}) - 1.04}$, where $\mu^* = 0.13$ is the Coulomb pseudopotential parameter, and substituting the relevant parameters yields the electro-acoustic coupling strength of $V_3Os_{1-2x}Si_xGe_x$ ($x = 0.333, 0.375, 0.425$) to be $\lambda_{ep} = 0.55$, 0.56, and 0.60, respectively, which suggests weakly coupled superconductors. From the electroacoustic coupling strength and the electron specific heat coefficient, the density of states at the Fermi energy level $D(E_F)$ can be calculated as follows: $D(E_F) = \dfrac{3\gamma}{\pi^2 k_B^2 (1+\lambda_{ep})}$. We obtain that the density of states of the sample $V_3Os_{1-2x}Si_xGe_x$ ($x = 0.333, 0.375, 0.425$) at the Fermi energy level is $D(E_F) = 5.32$ states/eV-f.u, 6.07 states/eV-f.u and 6.63 states/eV-f.u., respectively, demonstrating that the elevation of $T_c$ is primarily driven by the increase in electronic density of states, which outweighs the influence of lattice softening. From the superconducting parameters summarized in Table 1, it can be inferred that $V_3Os_{1-2x}Si_xGe_x$ ($x = 0.333$, 0.375, 0.425) are weak-coupled type-II s-wave superconductors.

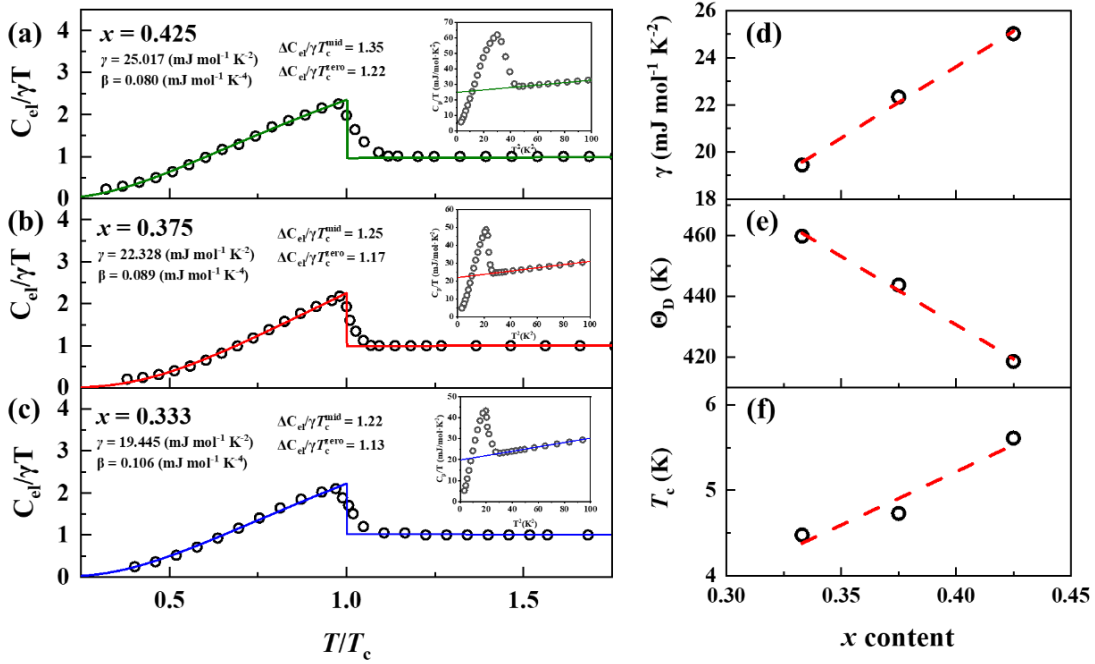

Figure 3 Normalized electronic specific heat $C_{el}/\gamma T_c$ as a function of reduced temperature $T/T_c$ for (a) $x = 0.425$, (b) $x = 0.375$, and (c) $x = 0.333$. The insets show the measured specific heat $C_p/T$ versus $T^2$. The lines in the insets are fit to $C_p = \gamma T + \beta T^3$ for $T > T_c$. (d) The $x$ content dependence of electronic specific-heat coefficients ($\gamma$), (e) the Debye temperature ($\Theta_D$) obtained from low-temperature fits of specific heats, and (f) $T_c$ of $V_3Os_{1-2x}Si_xGe_x$ ($x = 0.333, 0.375, 0.425$) MEAs.

The critical current density ($J_c$) of the $V_3Os_{1-2x}Si_xGe_x$ ($x = 0.333, 0.375, 0.425$) MEAs was determined from the magnetization hysteresis (M-H) loops employing the Bean model[41], expressed as $J_c = 20\Delta M/[a(1 - \frac{a}{3b})]$, where ΔM is the magnetic moment difference between the upper and lower branches of the loop, and a and b (a < b) represent the sample cross-sectional dimensions. Fig. 4a shows the magnetic hysteresis loop and the semi-logarithmic plot of $V_3Os_{1-2x}Si_xGe_x$ ($x = 0.333, 0.375, 0.425$) of the isothermal $J_c$ at 2 K and 3 K, respectively. The zero-field $J_c$ values for $x = 0.333$ are evaluated to be approximately $1.48 \times 10^6$ A/cm$^2$ at 2 K and $9.07 \times 10^5$ A/cm$^2$ at 3 K. For $x = 0.375$, the $J_c$ is $8.48 \times 10^6$ A/cm$^2$ at 2 K and $2.93 \times 10^6$ A/cm$^2$ at 3 K. In terms of $x = 0.425$, the $J_c$ is $3.72 \times 10^6$ A/cm$^2$ at 2 K and $2.77 \times 10^6$ A/cm$^2$ at 3 K. There are limited reports on the critical current density of A15-type MEAs or HEAs at low temperatures (2-3 K). Consequently, we compared our results with the $J_c$ data of other MEAs and HEAs measured at similar temperatures. Nikita Sharma et al. synthesized a $Nb_{0.25}Ta_{0.25}Ti_{0.25}Zr_{0.25}$, which exhibits a $J_c$ of $10^5$ A/cm$^2$ at 2 K and 3 K, despite possessing a $T_c$ of 8.0 K[42]. Similarly, The zero-field $J_c$ of $Ta_{1/6}Nb_{2/6}Hf_{1/6}Zr_{1/6}Ti_{1/6}$ HEA was reported to be $10^4$ A/cm$^2$ at 2 K[43]. Notably, all V$_3$Os$_{1-2x}$Si$_x$Ge$_x$ ($x$ = 0.333, 0.375, 0.425) MEAs synthesized in this work exceed these reported values. Moreover, our samples surpass the recognized practical benchmark of $10^5$ A/cm$^2$ required for high-field superconducting magnet technologies, demonstrating their superior potential for industrial applications[44].

The field dependence of $J_c$ is analyzed using the Kramer model, where $J_c^{0.5}H^{0.25}$ is plotted against $H$ to determine the irreversibility field ($H^*$) via linear extrapolation, as shown in Fig. 4b. The red lines indicate that the high-field data follow a linear trend consistent with flux-line lattice shear pinning. As the $x$ increases from 0.333 to 0.425, a significant enhancement in $H^*$ is observed, reaching a maximum of 3.3 $T$ for $x$ = 0.425 at 2 K. The pronounced upturn at low fields suggests a transition from individual vortex pinning or weak-link dominated behavior to a collective pinning regime at higher fields.

The normalized pinning force density $f_P$ (= $F_P/F_{P,max}$) is plotted as a function of the reduced magnetic field $h (= H/H^*)$ to investigate the scaling behavior and dominant pinning mechanisms across different compositions $x$ and temperatures, as shown in Fig. 4c. The data for all samples collapse onto a single scaling curve, indicating a unified pinning mechanism independent of temperature and stoichiometry within this range. The peak position $h_{max}$ occurs at approximately 0.2, which is characteristic of strong surface pinning (Higuchi model) rather than point pinning ($h_{max} \approx 0.33$). The solid black line and the blue dashed line represent the fit for the strong (surface) pinning model of DewHuges, $f_P \propto h^{0.5}(1-h)^2$ [45], and a double exponential model using $f_P(h) = J_1\exp(-B_1h)h + J_2\exp(-B_2h)h$[46], respectively. Both of which describe the data well, confirming that grain boundaries or planar defects likely act as the primary flux pinning centers in V$_3$Os$_{1-2x}$Si$_x$Ge$_x$ ($x$ = 0.333, 0.375, 0.425) MEAs.

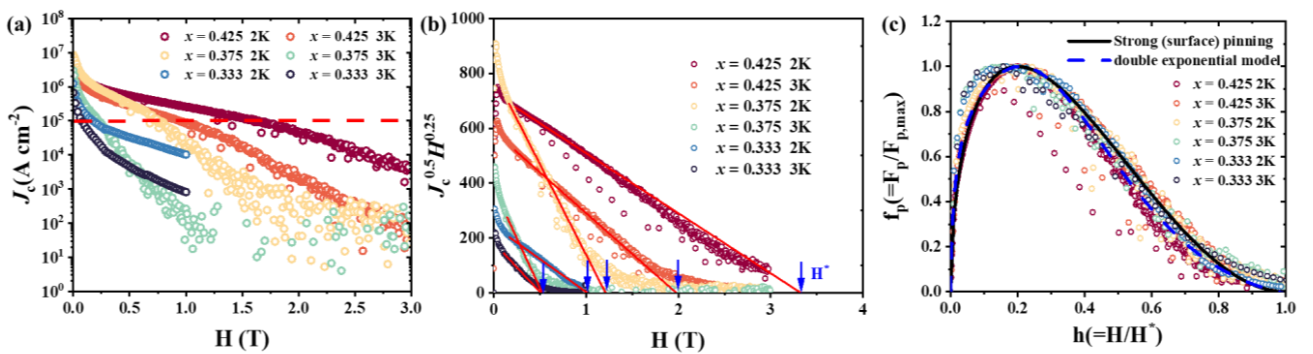

Figure 4 (a) Semi-log plot of magnetic field dependence of the critical current density $J_c$ of V$_3$Os$_{1-}$

$_{2x}Si_xGe_x$ ($x$ = 0.333, 0.375, 0.425) at 2 K and 3 K. (b) Kramer plots (solid lines) to determine the irreversible fields $H^*$. (c) Normalized flux pinning force density $f_P$ (= $F_P/F_{P, max}$) with the reduced field $h$(= $H/H^*$) of the $V_3Os_{1-2x}Si_xGe_x$ ($x$ = 0.333, 0.375, 0.425) at 2 K and 3 K. The solid line and the dashed line are fitting curves obtained for the double exponential model and the surface pinning mechanisms model, respectively.

Table 1 Superconducting parameters for the $V_3Os_{1-2x}Si_xGe_x$ MEAs and compared with binary $V_3Si$ and $V_3Ge$.

| | $x$ = 0.333 | $x$ = 0.375 | $x$ = 0.425 | $V_3Si$ | $V_3Ge$ |
|---|---|---|---|---|---|
| $\Delta S_{mix}$ | 0.837 | 0.833 | 0.815 | – | – |
| $T_c$ (K) | 4.48 | 4.73 | 5.61 | 16.3 ~ 17 | 6 ~ 6.3 |
| RRR | 1.09 | 1.05 | 1.13 | 8.97 ~ 22 | 5.2 |
| $\mu_0 H_{c1}$ (mT) | 8.36 | 7.13 | 8.54 | 122 ~ 138 | 56 |
| $\mu_0 H_{c2}$ (T)$^{GL}$ | 8.97 | 7.68 | 8.86 | 30.57 | 4 ~ 5.8 |
| $\mu_0 H_{c2}$ (T)$^{Pauli}$ | 8.33 | 8.80 | 10.43 | 30.318 ~ 31.62 | 11.16 ~ 11.72 |
| $\gamma$ (mJ mol$^{-1}$ K$^{-2}$) | 19.445 | 22.328 | 25.017 | 55.4 ~ 72.1 | 31.2 |
| $\beta$ (mJ mol$^{-1}$ K$^{-4}$) | 0.106 | 0.089 | 0.080 | 0.03 | 0.098 |
| $\Delta C_{el}/\gamma T_c$ | 1.22 | 1.25 | 1.35 | 1.81 ~ 2.09 | 1.33 |
| $\Theta_D$ (K) | 459.8 | 443.7 | 418.6 | 420 | 430 |
| $2\Delta_0/k_B T_c$ | 3.26 | 3.30 | 3.43 | 3.54 | 3.42 |
| $\lambda_{ep}$ | 0.55 | 0.56 | 0.60 | 1.0 ~ 1.1 | 0.61 |
| $J_c$(A cm$^{-2}$) | 1.24×10$^6$ (2 K) 6.64×10$^5$ (3 K) | 8.48×10$^6$ (2 K) 2.93×10$^6$ (3 K) | 3.72×10$^6$ (2K) 2.77×10$^6$ (3K) | < 8.7 × 10$^5$ (film, 5 - 15 K),N/A 2×10$^7$ (10 K) | |
| References | This work | This work | This work | [47-50] | [16, 17] |

Fig. 5 systematically demonstrates the correlation analysis between the $T_c$ and Valence Electron Concentration (VEC) for A15-type binary $V_3Os$, $V_3Si$, $V_3Os_{1-2x}Si_xGe_x$ ($x$ = 0.333, 0.375, 0.425), $V_3M$ alloys (M = Ir, Pd, Au, Al, Ga, Sn)[48, 51-54]. The parent compounds $V_3Si$ and $V_3Ge$ exhibit superconducting transitions at 17 K and 6 K, respectively, while $V_3Os$ remains non-superconducting. For the medium-entropy alloy $V_3(Os_{0.333}Si_{0.333}Ge_{0.333})$, a reduced $T_c$ of 4.48 K is observed. However, $T_c$ undergoes a systematic recovery upon decreasing the Os content, rising to 4.73 K and 5.61 K as the Os atomic fraction is reduced to 0.25 and 0.15, respectively. This variation in $T_c$ correlates strongly with the VEC, suggesting that the density of states tailored by Si/Ge doping is the primary driver for $T_c$ enhancement, consistent with the Matthias empirical rule. As Os concentration declines, the VEC shifts from 5.08 toward 4.9, approaching the critical threshold of e/a ≈ 4.7 where $T_c$ typically peaks in transition metal-based superconductors. In addition to the electronic tuning described by the Matthias rule, the observed enhancement in $T_c$ can also be interpreted through the "cocktail effect," a characteristic feature of HEAs/MEAs where the final properties are a synergistic reflection of their constituents [55,56]. Given that the parent compounds $V_3Si$ ($T_c$ ≈ 17 K) and $V_3Ge$ ($T_c$ ≈ 6 K) possess significantly higher $T_c$ than the non-superconducting $V_3Os$, the incremental substitution of Os with Si and Ge naturally elevates the $T_c$ of the resulting MEAs. This complementary perspective suggests that the superconductivity in $V_3(Os_{1-2x}Si_xGe_x)$ ($x$ = 0.333, 0.375, 0.425) is governed by both the global electronic environment and the beneficial intrinsic properties of its high-$T_c$ binary components. On the other hand, the high configurational entropy is essential for stabilizing the cubic A15 phase against phase segregation and introducing chemical disorder that enhances electron scattering, which is beneficial for the upper critical field[57,58].

Detailed doping mechanism studies show that Ga/Al substitution at silicon sites reduces the e/a ratio (as Ga/Al exhibits lower electronic contributions than Si), while Co doping at vanadium sites increases e/a through the introduction of higher-valent elements. However, amorphous systems exhibit significantly broader extremal regions compared to simple crystalline superconductors. The $T_c$ - e/a trajectory of the $V_3Si$ system not only matches the variation amplitude of crystalline alloys but also shows good consistency with the monotonic trend observed in amorphous superconductors. Specifically, $V_3Si$ achieves its maximum $T_c$ at VEC = 4.7, a characteristic feature that validates the core prediction of the Matthias rule while revealing inherent limitations in enhancing $T_c$ values of $V_3Si$ superconducting materials through chemical doping. So far, many chemical doping studies have been conducted to enhance the superconductivity of $V_3Si$ [54,59,60]. However, the effect of doping-induced changes in the VEC on $T_c$ is still worthy of further study.

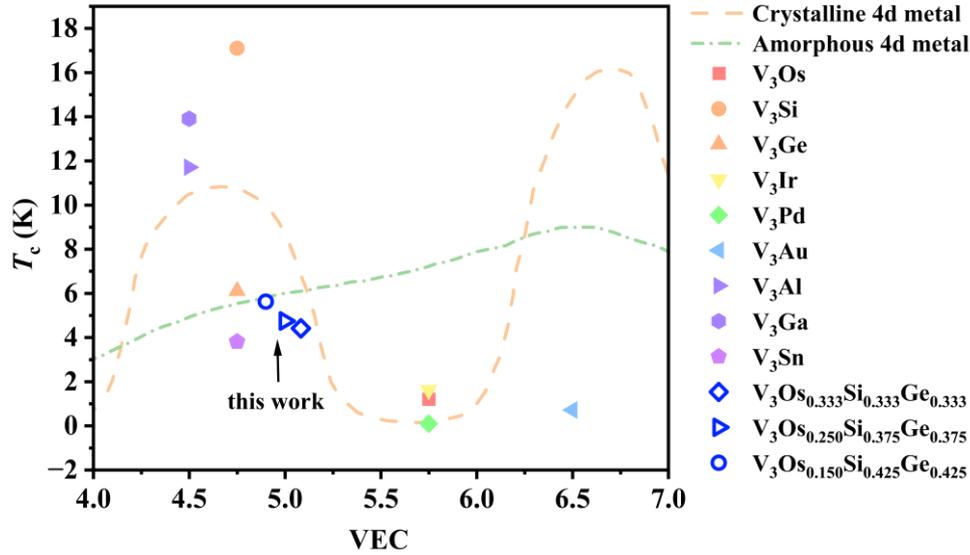

Figure 5 The dependence of VEC on temperature in A15-structured superconductors.

## CONCLUSIONS

In summary, we successfully synthesized the A15-type $V_3Os_{1-2x}Si_xGe_x$ ($x$ = 0.333, 0.375, 0.425) MEAs via an arc melting method. Resistivity, magnetic susceptibility, and specific heat data confirm their identity as a type-II superconductor with a $T_c$ of 4.48 K, 4.73 K, and 5.61 K, respectively. For the $x$ = 0.333 sample, the relatively high upper critical field reaches 8.97 T, which exceeds the Pauli paramagnetic limiting field of 8.33 T. This strong performance is attributed to the significant suppression of paramagnetic pair-breaking (likely due to the strong spin-orbit coupling induced by Os substitution), positioning the alloy as a promising platform for exploring high-field superconductivity.

Furthermore, zero-field critical current density, $J_c$, values demonstrated exceptional performance. All $V_3Os_{1-2x}Si_xGe_x$ ($x$ = 0.333, 0.375, 0.425) MEAs exceed the $10^5$ A/cm$^2$ practical benchmark, underscoring their potential for demanding technological applications. These results highlight the efficacy of entropy engineering in modulating the electronic structure of A15-type compounds, providing a strategic pathway for the discovery of next-generation high-performance superconductors.

**Acknowledgements**

This work is supported by the Natural Science Foundation of China (No. 12274471, 12404165), Guangdong Provincial Science and technology plan project - International Science and technology

cooperation field (No. 2025A0505020045), Guangdong Major Project of Basic Research (2025B0303000004), Guangdong Basic and Applied Basic Research Foundation (No. 2025A1515010311), Guangzhou Science and Technology Programme (No. 2024A04J6415), the Open Research Fund of State Key Laboratory of Quantum Functional Materials (NO. QFM2025KF004), the State Key Laboratory of Optoelectronic Materials and Technologies (Sun Yat-Sen University, No. OEMT-2024-ZRC-02), and Key Laboratory of Magnetoelectric Physics and Devices of Guangdong Province (Grant No. 2022B1212010008), and Research Center for Magnetoelectric Physics of Guangdong Province (2024B0303390001). Lingyong Zeng was thankful for the Postdoctoral Fellowship Program of CPSF (GZC20233299) and the Fundamental Research Funds for the Central Universities, Sun Yat-sen University (29000-31610058).

# Supporting Information

# Superconductivity in the A15-type V$_3$(Os$_{1-2x}$Si$_x$Ge$_x$) medium-entropy alloys

*Yucheng Li[1,#], Kuan Li[1,#], Lingyong Zeng[1], Rui Chen[1], Jingjun Qin[1], Shuangyue Wang[1], Huixia Luo[1,2,,3,4*]*

[1] School of Materials Science and Engineering, Sun Yat-sen University, No. 135, Xingang Xi Road, Guangzhou, 510275, P. R. China

[2] State Key Laboratory of Optoelectronic Materials and Technologies, Sun Yat-sen University, No. 135, Xingang Xi Road, Guangzhou, 510275, P. R. China

[3] Key Lab of Polymer Composite & Functional Materials, Sun Yat-sen University, No. 135, Xingang Xi Road, Guangzhou, 510275, P. R. China

[4] Guangdong Provincial Key Laboratory of Magnetoelectric Physics and Devices, Sun Yat-sen University, No. 135, Xingang Xi Road, Guangzhou, 510275, P. R. China

[#] These authors contributed equally to this work.

*Corresponding author/authors complete details (Telephone; E-mail:) (+86)-2039386124; E-mail address: luohx7@mail.sysu.edu.cn;

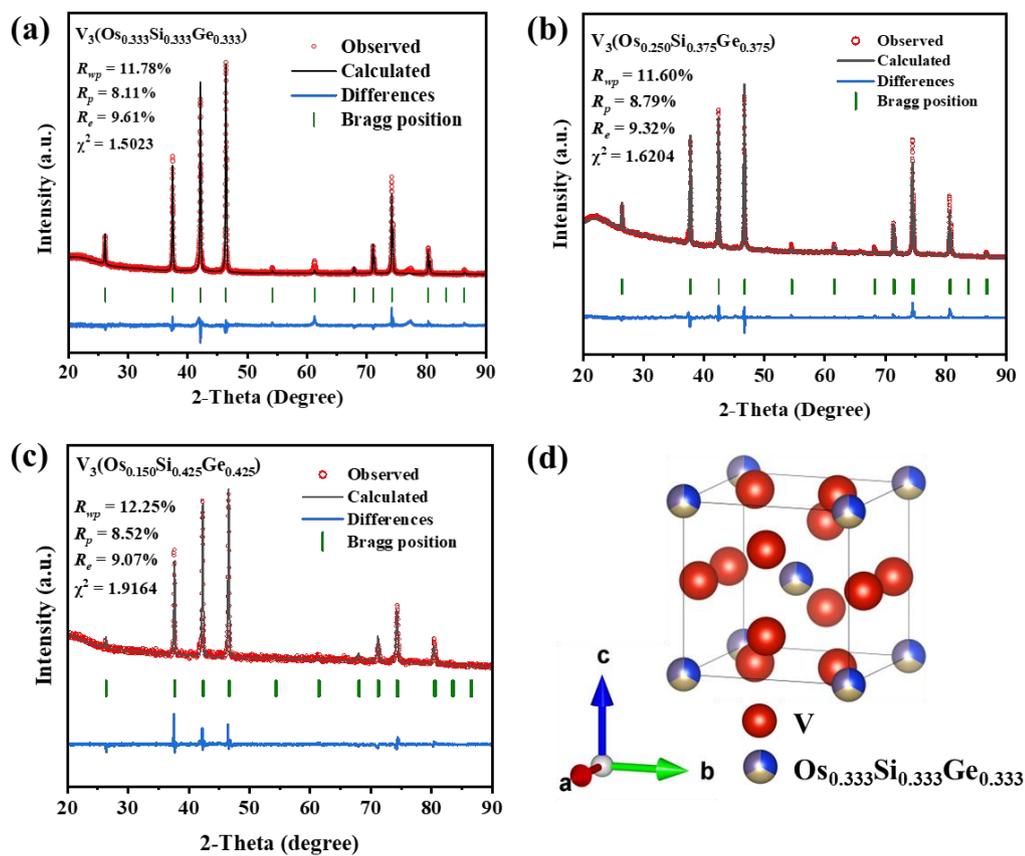

Figure S1(a-c) The Rietveld refinement profiles of the XRD patterns for $V_3Os_{1-2x}Si_xGe_x$ ($x$ = 0.333, 0.375, 0.425) samples. (d) The crystal structure for $V_3Os_{1-2x}Si_xGe_x$ ($x$ = 0.333) sample.

Table S1 Refined factors of $V_3Os_{1-2x}Si_xGe_x$ ($x$ = 0.333, 0.375, 0.425) MEAs.

| | $x$ = 0.333 | $x$ = 0.375 | $x$ = 0.425 |
|---|---|---|---|
| Space Group | \multicolumn{3}{c}{$Pm\bar{3}n$ (No. 223)} | | |
| Lattice parameter $a$ (Å) | 4.854 | 4.805 | 4.764 |
| $R_{wp}$ (%) | 11.78 | 11.60 | 12.25 |
| $R_p$ (%) | 8.11 | 8.79 | 8.52 |
| $R_e$ (%) | 9.61 | 9.32 | 9.07 |
| $\chi^2$ | 1.5023 | 1.5204 | 1.9164 |
| V (6$c$ site) | (0.25, 0.5, 0) | | |
| Os/Si/Ge (2$a$ site) | (0, 0, 0) | | |
| Occupancy | V: 1.0<br>Os: 0.333<br>Si: 0.333<br>Ge: 0.333 | V: 1.0<br>Os: 0.250<br>Si: 0.375<br>Ge: 0.375 | V: 1.0<br>Os: 0.150<br>Si: 0.425<br>Ge: 0.425 |

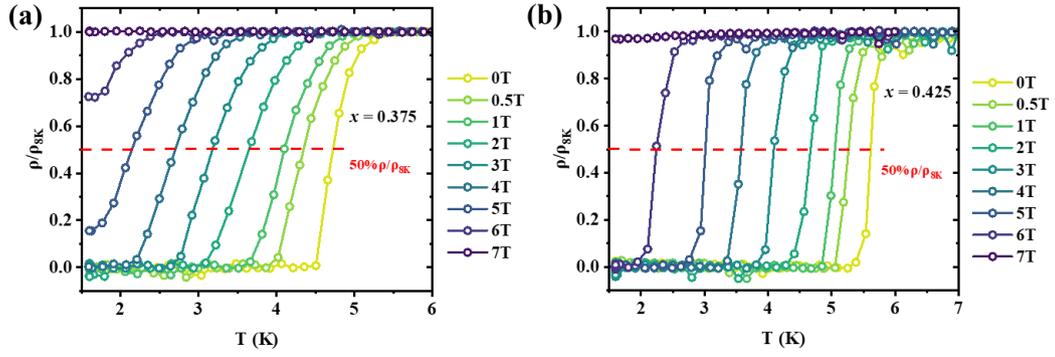

Figure S2. The resistivity transition at the field of 0-7 T for (a) $V_3Os_{0.250}Si_{0.375}Ge_{0.375}$ and (b) $V_3Os_{0.150}Si_{0.425}Ge_{0.425}$.

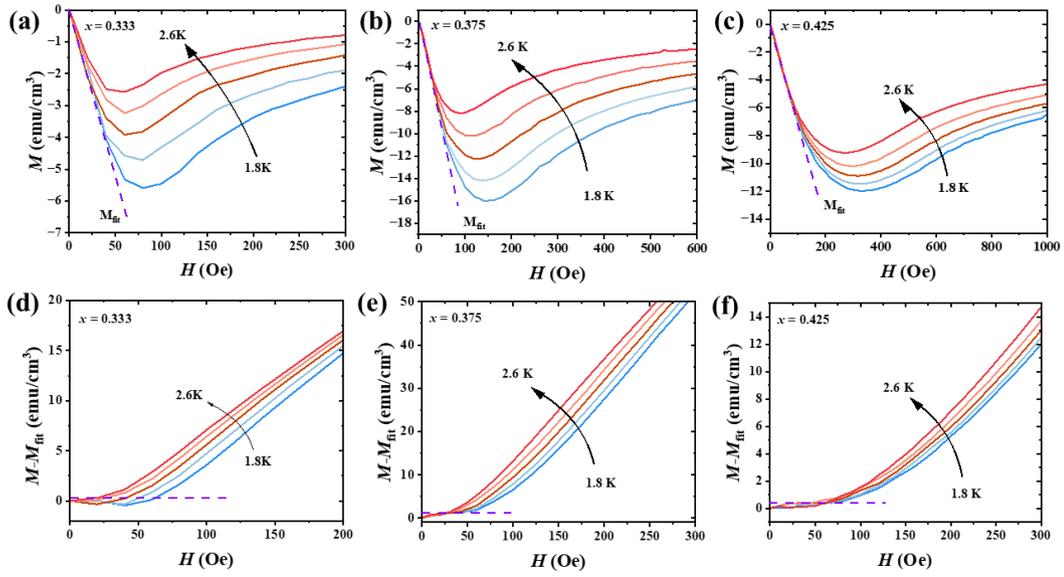

Figure S3. (a-c) The *M-H* curves at 1.8-2.6 K for $x$ = 0.333, 0.375, 0.425, respectively. (d-f) The $M$–$M_{fit}$ as a function of the applied field for $x$ = 0.333, 0.375, 0.425, respectively.